\documentclass[final,3p,times]{elsarticle}
\usepackage{amssymb}
\usepackage{amsmath}
\journal{The 17th annual conference of the International Association for Mathematical Geosciences,
September 5-13, 2015, Freiberg (Saxony) Germany \hspace{2cm}}

\begin{document}
\begin{frontmatter}
\title{Meshless RBF based pseudospectral solution of acoustic wave equation}
\author[label1]{Pankaj K Mishra}
\address[label1]{Department of Geology and Geophyiscs, Indian Institute of Technology, Kharagpur, India}
\ead{pankajkmishra01@gmail.com}
\ead[url]{sites.google.com/site/pankajkmishra01}

\author[label1]{Sankar K Nath\corref{cor1}}
\cortext[cor1]{Corresponding Author}
\ead{nath@gg.iitkgp.ernet.in}

\begin{abstract}
Chebyshev pseudospectral (PS) methods are reported to provide highly accurate solution using polynomial approximation. Use of polynomial basis functions in PS algorithms limits the formulation to univariate systems constraining it to tensor product grids for multi-dimensions. Recent studies have shown that replacing the polynomial by radial basis functions in pseudospectral method (RBF-PS) has the advantage of using irregular grids for multivariate systems. A RBF-PS algorithm has been presented here for the numerical solution of inhomogeneous Helmholtz's equation using Gaussian RBF for derivative approximation. Efficacy of RBF approximated derivatives has been checked  through error analysis comparison with PS method. Comparative study of PS, RBF-PS and finite difference approach for the solution of a linear boundary value problem has been performed. Finally, a typical frequency domain acoustic wave propagation problem has been solved using Dirichlet boundary condition and a point source. The algorithm presented here can be extended further for seismic modeling with complexities associated with absorbing boundary conditions. 

\end{abstract}
\end{frontmatter}
\section{Introduction}

In recent years, there have been significant developments in the meshless approaches for solving differential equations arising in various disciplines of science and engineering. Many formulations have been proposed for meshless solutions of differential equations \cite{Kansa19901,Dehg2007,NG2008} out of which, approaches based on radial basis functions are found to be efficient for modeling physical phenomena like structural mechanics \cite{Ferr2006}, fluid mechanics \cite{Demi2008} etc. Pseudospectral methods, also known as discrete variable representation (DVR) methods are highly accurate approaches \cite{Hoss2005,Fass2005,Fass2007} incorporating the idea of approximating the field using a set of smooth and global basis function like polynomials. Use of polynomial basis functions makes the formulation univariate constraining the algorithm limited to tensor product grids for higher dimensions. An improved pseudospectral algorithm was proposed replacing polynomials with radial basis functions making the formulation to incorporate irregular girding as well as irregular geometries. Such approach is collectively termed as radial basis based pseudospectral method (RBF-PS)\cite{Fass2007,MU2012}.

Here, we present a RBF-PS algorithm for numerical solution of inhomogeneous Helmholtz's equation which is also a frequency domain representation of acoustic wave equation. An accuracy analysis of RBF approximated derivatives has been performed through a numerical test involving a comparative study of maximum error in derivative approximation with those obtained using polynomial approximation over Chebyshev differential grids. For the reason that pseudospectral methods are explained as high accuracy limit of finite difference method \cite{Forn1996}, a comparative test has been performed among finite difference, PS and RBF-PS approaches by solving a linear boundary value problem. Finally, the algorithm has been used for solving two dimensional frequency-domain representation of acoustic wave equation using differential Chebyshev grids.
\section{Derivative approximation}
\noindent RBF-PS method involves the common approach of all weighted residual methods i.e. approximating the field using combination of some pre-defined functions termed as shape functions and some unknown coefficients which are determined depending upon the governing equation of the problem. In RBF-PS method, shape functions are computed using radial basis functions. The approximate solution can be expressed as,

\begin{eqnarray} 
p^{a}(\mathbf x)= \sum\limits_{k=1}^N \gamma_k \phi_k (\mathbf x) 
\end{eqnarray}
\noindent or, 
\begin{eqnarray}
\mathbf p^{a} =   \mathbf \Theta \Gamma 
\end{eqnarray}

\noindent where the elements of matrix $\Theta$ are such that $\Theta_{i,j}=\phi_i(x_j)$, $\gamma_k = [ \gamma_1, \gamma_2, \gamma_3, ...\gamma_N]^T$ are the unknown coefficients, and $ \phi_k = [ \phi_1 (\mathbf x),\phi_2 (\mathbf x),\phi_3 (\mathbf x)...\phi_N (\mathbf x)]^T $ are set of shape functions. For shape function construction here, the Gaussian radial basis function has been considered. The expressions for general Gaussian radial basis function and its derivatives are given as,
\begin{eqnarray}
\Phi(r) = e^{ -(\epsilon r)^2} 
\end{eqnarray}
\begin{eqnarray}
\Phi^{'} (r) = -2\epsilon^2 r e^{ -(\epsilon r)^2}  
\end{eqnarray}

\begin{eqnarray}
\Phi^{''} (r) = 2\epsilon^2(2(\epsilon r)^2-1)e^{ -(\epsilon r)^2}
\end{eqnarray}

\noindent Taking derivative of equation (1), it can be written as, 
\begin{eqnarray} 
\frac{\partial}{\partial x} p^{a}(\mathbf x)= \sum\limits_{k=1}^N \gamma_k \frac{\partial}{\partial x}\phi_k (\mathbf x) 
\end{eqnarray}

\noindent or in parametric form we can write it as, 
\begin{eqnarray}
D [\mathbf p^{a}] =   \mathbf \Theta_x \Gamma 
\end{eqnarray}
\noindent Where $D$ denotes the first derivative operator and $\Theta_x$ contains first derivatives of RBFs. For univariate case invertibility of $\Theta_x$ is assured. This is one of the reason why pseudospectral methods are generally used for univariate system. In order to use equation (7) for derivative approximation, invertibility of $\Theta$ must be assured, which depends on the choice of the radial basis function and the grid arrangements. Using equation (2), equation(7) can be written as, 
\begin{eqnarray}
D [\mathbf p^{a}] =  \mathbf \Theta_x \Theta^{-1} \mathbf p^{a}
\end{eqnarray} 
The differential operator D, therefore can be expressed as, 
\begin{eqnarray}
D = \Theta_x \Theta^{-1}
\end{eqnarray}

\section{Discretization Scheme}
\noindent Let's consider a general form of our concerned governing equation in a domain $\Omega$ inside the boundary $\mathcal{B}$. 
\begin{eqnarray}
\nabla^2 p + k^2 p = S \qquad \in \Omega
\end{eqnarray}
Using the convention adopted in equation (1) we can write equation (10) as, 
\begin{eqnarray}
\mathcal{H} p=S \qquad \in \Omega
\end{eqnarray}
\noindent with Dirichlet boundary condition 
\begin{eqnarray}
p = g \qquad \in \mathcal{B}
\end{eqnarray}
using the similar convention as equation(9) for the Laplace operator, problem given in equations (11) and (12) can be written in parametric form as given by,
\begin{eqnarray}
\begin{bmatrix}
\Theta_\emph{L} \Theta^{-1} + k^2 \\
\Theta
\end{bmatrix}
\begin{bmatrix}
\mathbf{\Gamma}
\end{bmatrix}
=
\begin{bmatrix}
S \\
g
\end{bmatrix}
\end{eqnarray} 

\noindent or 
\begin{eqnarray}
\begin{bmatrix}
\mathbf{\Gamma}
\end{bmatrix}
=
\begin{bmatrix}
\Theta_\emph{L} \Theta^{-1} + k^2 \\
\Theta
\end{bmatrix}^{-1}
\begin{bmatrix}
S \\
g
\end{bmatrix}
\end{eqnarray}
In accordance with equation (1) the approximated field variable $\mathbf{p^a}$ can be written as,
\begin{eqnarray}
\begin{bmatrix}
\mathbf{p^a}
\end{bmatrix}
= \Theta
\begin{bmatrix}
\Theta_\emph{L} \Theta^{-1} + k^2 \\
\Theta
\end{bmatrix}^{-1}
\begin{bmatrix}
S \\
g
\end{bmatrix}
\end{eqnarray}
Thus the differential operator for the governing equation can be computed as
 $ \Theta^{-1}
\begin{bmatrix}
\Theta_\emph{L} \Theta^{-1} + k^2 \\
\Theta
\end{bmatrix}
$.
Here we use Chebyshev grids for discretization of domain and Kronecker delta product to compute the Helmholtz operator as given below, 
\begin{eqnarray}
\mathcal{H} = \mathbb{I} \otimes D^2 + D^2 \otimes \mathbb{I} + k^2
\end{eqnarray}
while using RBF approximation methods, role of shape parameter is very crucial. For this study, we have adopted the work of \cite{Fass20077} to determine the optimal shape parameter. 
\section{Numerical Tests}
Here we compute maximum error in derivative approximation of two functions using RBF approximation and compare the results with those obtained using Chebyshev pseudospectral differentiation as well as with exact differentiation as given by,
\begin{eqnarray}
f(x)= \parallel x^{3} \parallel \qquad f^{'}(x)= 3x\parallel x \parallel \qquad and,
\end{eqnarray}
\begin{eqnarray}
g(x) = \exp(-x^{-2}) \qquad g^{'}(x)= \frac{2}{x^3} \exp(-x^{-2})
\end{eqnarray}
%++++++++++++++++++++++++++++++++++++++++++++++++++++++++++++++++++++++
\begin{figure}[!htb]
\centering
\includegraphics[scale=0.4]{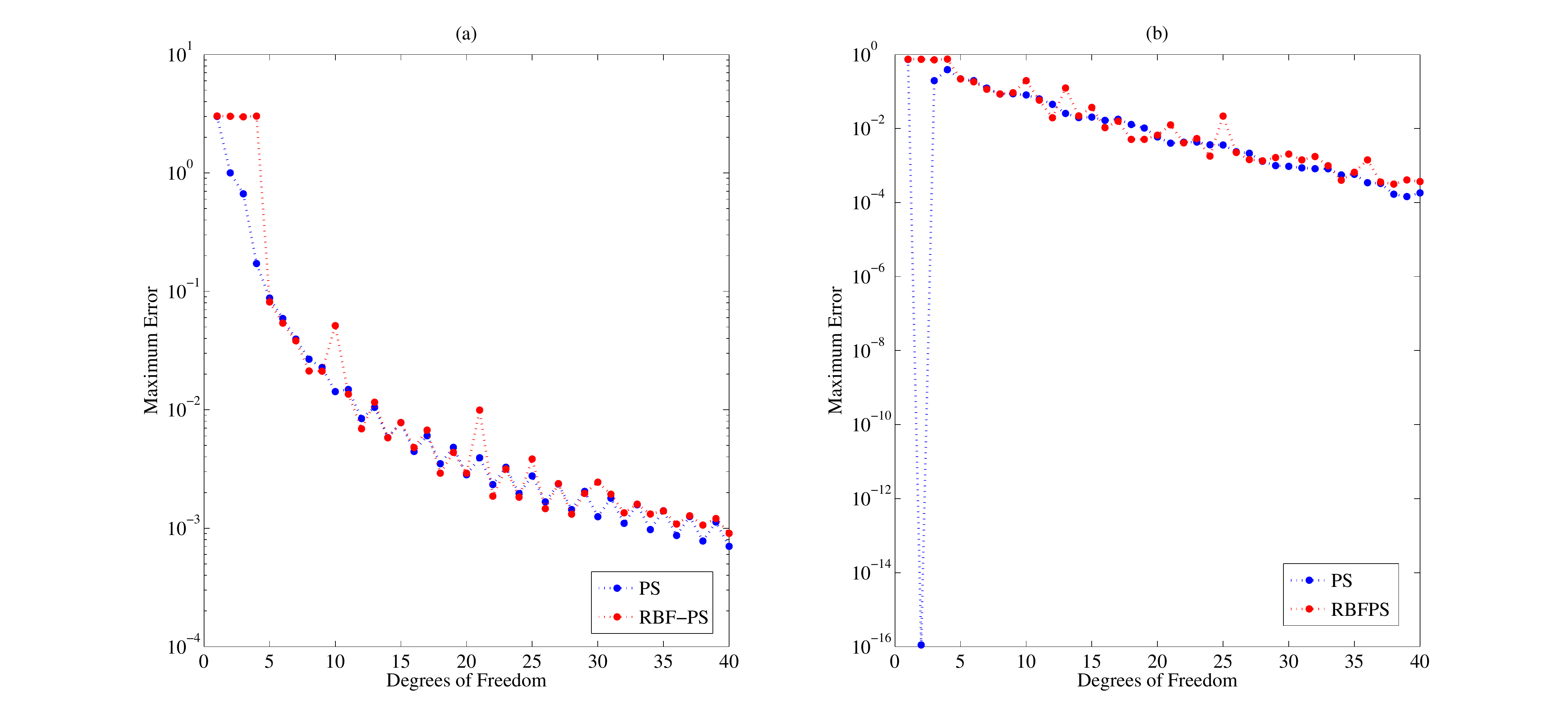}
\caption{Maximum error in derivative approximation via radial basis functions and its comparison with Chebyshev pseudospectral approximation}
\end{figure}
\begin{figure}[!htb]
\centering
\includegraphics[scale=0.4]{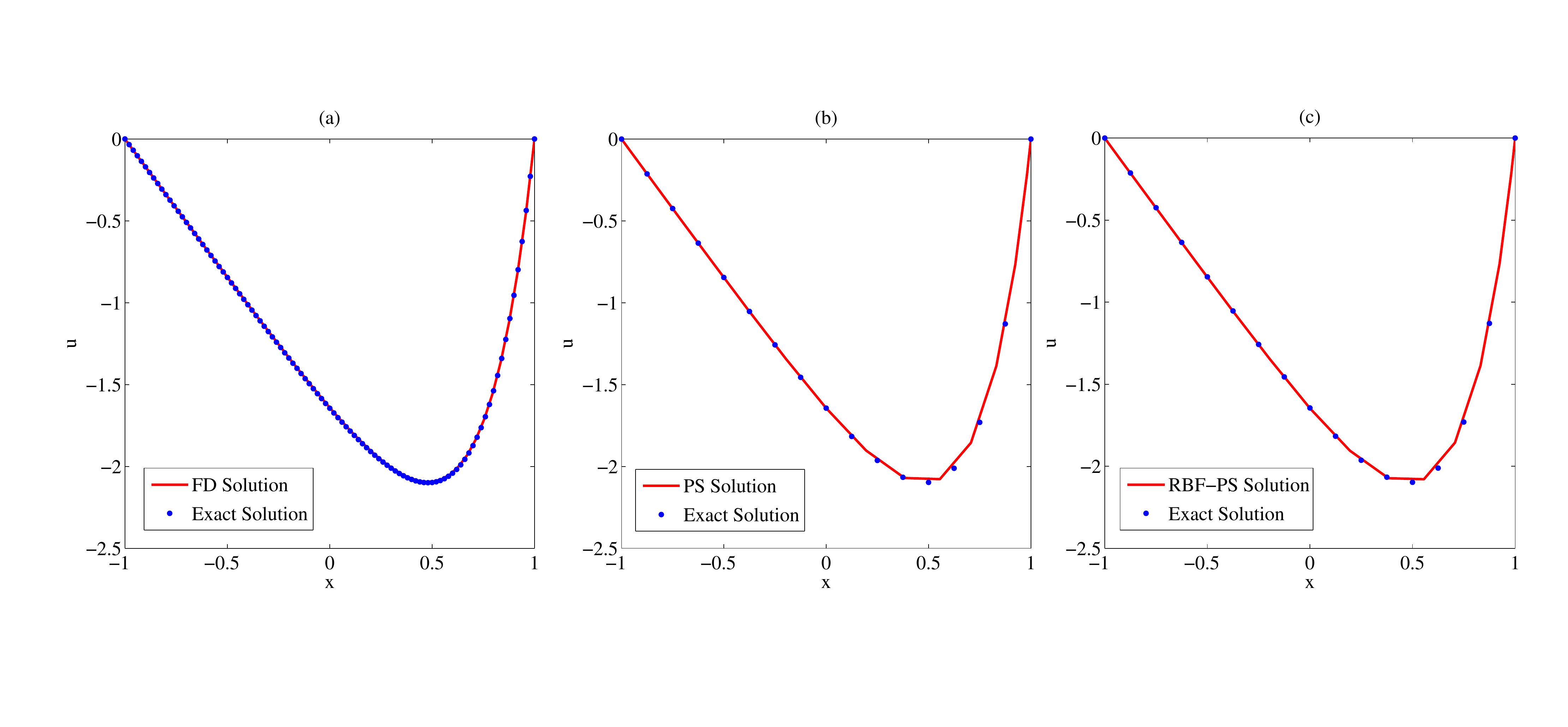}
\caption{Solution of the linear boundary value problem using (a) finite difference (b) Pseudospectral method (c) RBF-PS method and its comparison with the exact solution \vspace{2cm}}
\end{figure}
%+++++++++++++++++++++++++++++++++++++++++++++++++++++++++++++++++++
maximum error in computing derivative using PS and RBF-PS approach have been shown in figure 1. Although application of RBF in pseudospectral methods makes it flexible to model more complex physical processes, this numerical test shows that the accuracy of RBF approximated derivatives are less than those obtained by Chebyshev pseudospectral differentiation. For the second test, we consider a linear boundary value problem as given by, 
\begin{eqnarray}
u^{''}(x) = \exp(4x) \qquad \in [-1,1]
\end{eqnarray}
with the boundary condition
\begin{eqnarray}
u(-1)=u(1)=0
\end{eqnarray}
The problem (19) with boundary conditions (20) has been solved using three approaches namely finite difference (FD), pseudospectral method (PS) and RBF based pseudospectral method (RBF-PS) method. For FD, 100 grid points has been used whereas for PS and RBF-PS method, 16 scattered nodes are used. Figure 2 shows the efficacy of RBF-PS method in accordance with PS method. 

Frequency domain representation of acoustic wave equation is widely used for full waveform inversion and the precises modeling of seismic attenuation processes [\cite{FJ2005}]. A simplified problem of frequency domain acoustic wave propagation has been solved using null Dirichelet boundary conditions. The problem formulation is given by,
 \begin{eqnarray}
 \frac{\partial^{2} p}{\partial x^{2}} + \frac{\partial^{2} p}{\partial z^{2}} + \frac{\omega^2}{{v_{p}}^2} p = S(x,y,\omega)
\end{eqnarray} 
\begin{eqnarray}
  S(x,y,\omega) =  \exp ((x-x_0)^2+(z-z_0)^2) \left[\frac{2f^2}{\pi^2 f^{3}_{c}} \exp \left( - \frac{f^2}{\pi f^2} \right) \right]
\end{eqnarray}
where $f = {\omega}\diagup{2\pi}$ is the frequency and $f_c$ is related to cutoff frequency as described in \cite{MO2014}. For this simplified test, the values $f=20Hz$ and $v_{p}=1$ has been used. The problem is solved using 100 nodes using Chebyshev differential gridding in the domain $[-1,1] \times[-1,1]$. The grid arrangement and an upscaled solution is shown in the figure $3(a)$ and $3(b)$ respectively.
\begin{figure}[!htb]
\includegraphics[width=7.8cm,height=7cm]{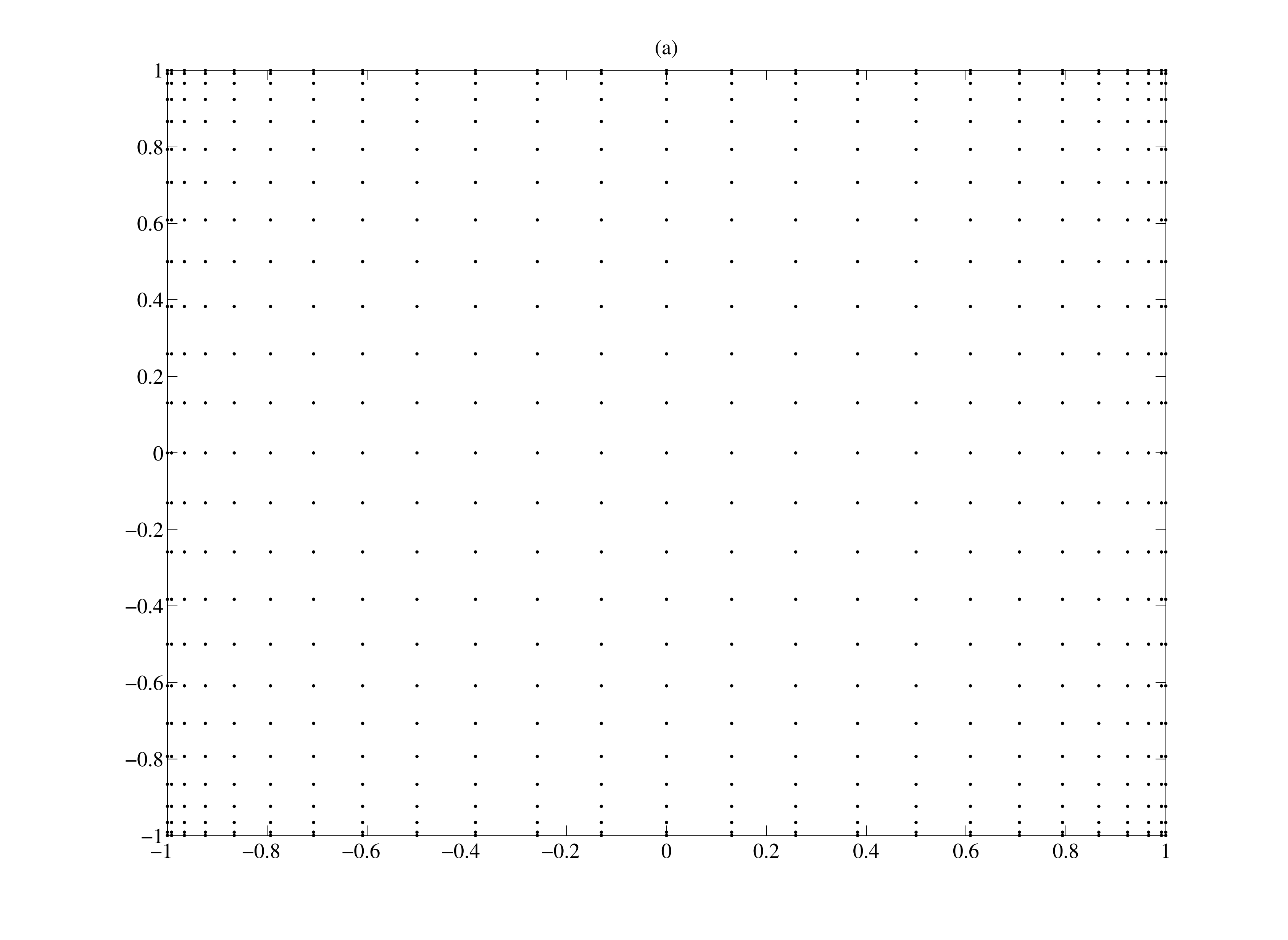}
\includegraphics[width=7.8cm,height=7cm]{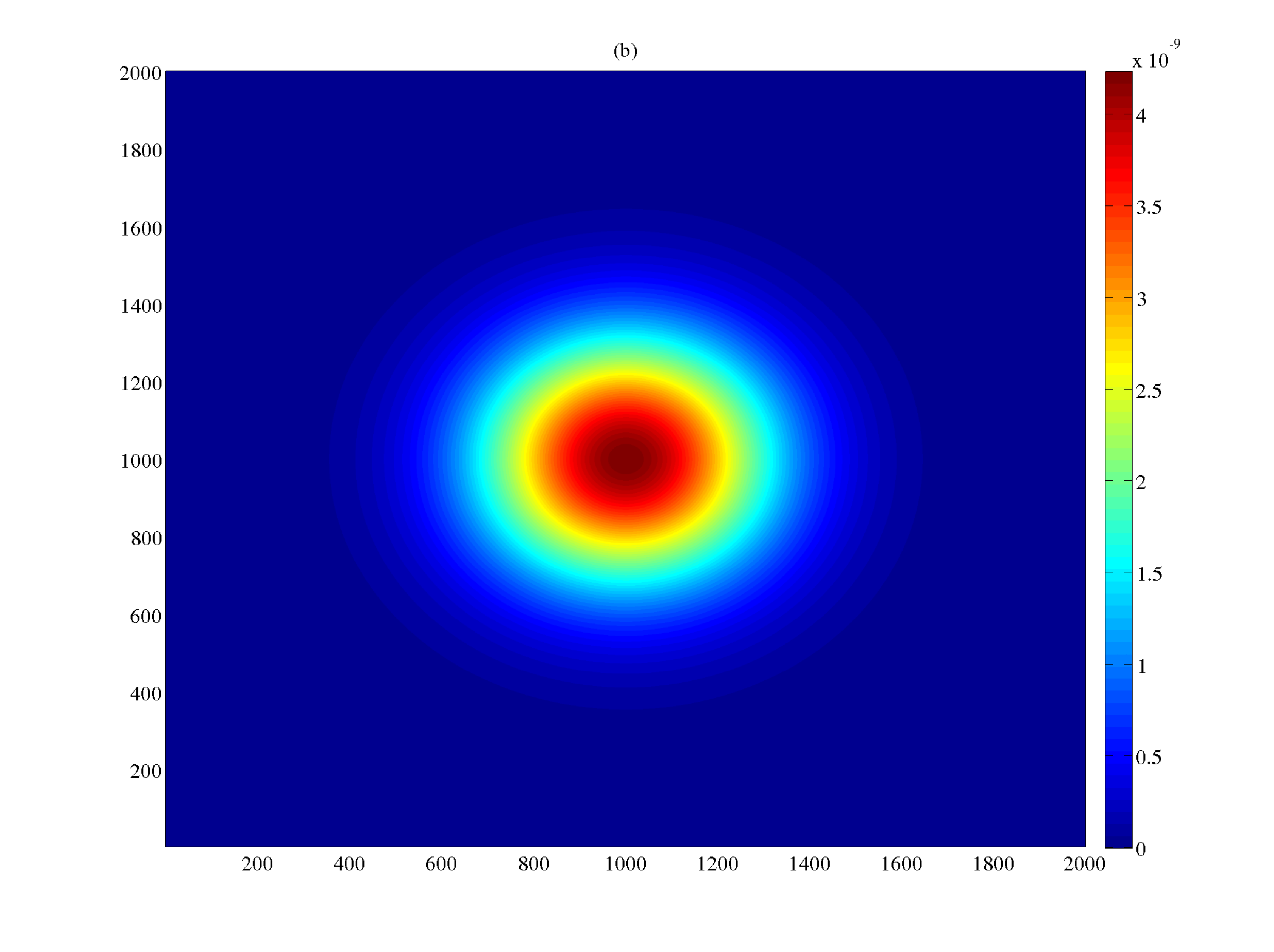}
\caption{(a) Chebyshev differential grids using 100 nodes in the domain and  (b) RBF-PS solution up-scaled to $2000\times2000$ grids exhibiting the high order smoothness of the solution.}
\end{figure}
%++++++++++++++++++++++++++++++++++++++++++++++++++++++++++++++++++++++
\section{Conclusion}
Though the accuracy of an RBF-PS algorithm suffers with the involvement of radial basis function, even then its accuracy is better than the conventional method of numerical solution. As the smoothness of the function increases, the accuracy of pseudospectral method increases while RBF-PS approach provides almost similar accuracy. With the application of appropriate absorbing boundary conditions RBF-PS algorithms can be used for precise seismic modeling and full waveform inversion.
\bibliographystyle{chicago}   % name your BibTeX data base

\end{document}